\documentclass[11pt,a4paper]{article}
\usepackage{jheppub}
\usepackage{amsmath,amssymb}
\usepackage{slashed}
\usepackage{graphicx}
\usepackage{subfigure}
\usepackage{mathptmx}
\usepackage{url}
\usepackage{pstricks}
\newcommand{\beq}{\begin{equation}}
\newcommand{\eeq}{\end{equation}}

\newcommand{\be}{\begin{eqnarray}}
\newcommand{\ee}{\end{eqnarray}}
\long\def\hidestart#1\hideend{}
\title
{Topological susceptibility in lattice Yang-Mills theory with open boundary
condition}
\author{Abhishek Chowdhury$^{a}$,}
\author{A. Harindranath$^{a}$,}
\author{Jyotirmoy Maiti$^{b}$ and}
\author{Pushan Majumdar$^{c}$ }

\affiliation{$^{a}$Theory Division, Saha Institute of Nuclear Physics \\
 1/AF Bidhan Nagar, Kolkata 700064, India}

\affiliation{$^{b}$Department of Physics, Barasat Government College,\\
10 KNC Road, Barasat, Kolkata 700124, India}

\affiliation{$^{c}$Department of Theoretical Physics, \\
Indian Association for
the Cultivation of Science, Kolkata 700032, India}

\emailAdd{abhishek.chowdhury@saha.ac.in}
\emailAdd{a.harindranath@saha.ac.in}
\emailAdd{jyotirmoy.maiti@gmail.com}
\emailAdd{tppm@iacs.res.in}
\date{November 19, 2013}
\abstract {We find that using open boundary condition in the temporal direction 
can yield the expected value of the topological susceptibility in lattice 
SU(3) Yang-Mills theory. As a further check, we show that the result agrees 
with numerical simulations employing the periodic boundary condition. 
Our results support the preferability of the open boundary condition over
the periodic boundary condition as the former allows for computation 
at smaller lattice spacings needed for continuum extrapolation at a lower 
computational cost.}
\begin{document}
\maketitle

\section{Motivation}
An open problem in numerical simulation of lattice QCD is that sampling gauge configurations over different topological sectors
becomes more and more difficult 
as the continuum limit is approached.
As a consequence, autocorrelation times of 
physical quantities grow rapidly making the calculation of expectation values 
time consuming. To partially 
overcome this problem, using open boundary conditions
(instead of the usual periodic or anti-periodic ones) in the temporal direction of the 
lattice has been proposed \cite{open0}. Lattice gauge theory with such boundary conditions have no 
barriers between different topological sectors. This has been shown
by extensive simulations in SU(3) gauge theory \cite{open1}. Even though the open 
boundary conditions introduce boundary effects and thus complicate the physics analysis, their advantage 
from the point of view of ergodicity and efficiency have been addressed in
simulations of 2+1 flavours of $O(a)$ improved Wilson quarks \cite{open2}. 
Advantages of 
using open boundary conditions have also been studied in the investigation of
SU(2) lattice gauge theory at weak coupling \cite{grady}. 

In the context of topology of gauge fields, an interesting quantity to study is the topological
susceptibility ($\chi$) in pure Yang-Mills theory  which is related to the $\eta^\prime$ mass
by the famous Witten-Veneziano formula \cite{witten,veneziano,seiler}. For recent high precision 
calculations of $\chi$ with periodic boundary condition see, for example, 
Refs. \cite{del,durr,lspb}.
Ref. \cite{del} uses Ginsparg-Wilson fermion for the topological charge density operator whereas
Ref. \cite{durr} uses the algebraic definition based on field strength tensor. 
A proposal to overcome the problem of short distance singularity in the
computation of topological susceptibility is given in Refs. \cite{lus2004,gl}. 
Ref. \cite{lspb} employs a spectral-projector formula which is designed to be free
from singularity and compares the result with that using the algebraic definition. The results 
using different approaches are in agreement with each other within statistical
uncertainties.

In this work we address the question whether an open boundary condition in the temporal direction can
yield the expected value of the topological susceptibility in SU(3) Yang-Mills theory.
We employ the algebraic definition for the topological charge density used in Ref.\cite{lspb}
and for a meaningful comparison with Ref.\cite{lspb} Wilson flow is used
to smoothen the gauge field. We also perform   
simulations with
periodic boundary conditions. We find that using an open boundary condition is  
advantageous as it allows one to sample different topological sectors by 
removing the barrier between them.

Unlike the periodic lattice, any physical quantity 
measured on a lattice with open boundary also has the additional boundary term along with the bulk part
(see for example Ref. \cite{as}).
In a simulation with all other parameters kept identical, the difference between the results for some 
physical quantity measured on a finite volume system with open and periodic boundary gives the boundary 
contribution for the system with the open boundary. As this boundary contribution diminishes with increasing 
volume, result from a system with open boundary approaches the same from a system with periodic boundary conditions.

 \begin{table}
\begin{center}
\begin{tabular}{|c|l|l|l|l|l|l|l|}
\hline \hline
Lattice & Volume & $\beta$ & $N_{\rm cnfg}$ &$N_0$ &$\tau$&$a[{\rm fm}]$ & $t_{0}/a^2$\\
\hline\hline
{$O_1$}&{$24^3\times48$}&{6.21} &{3970} &{12} &{3}& {0.0667(5)} & {6.207(15)}\\
\hline
{$O_2$}&{$32^3\times64$}&{6.42} &{3028} &{20}&{4}&{0.0500(4)} & {11.228(31)}\\
\hline
{$O_3$}&{$48^3\times96$}&{6.59} &{2333} &{26}&{5}&{0.0402(3)} & {17.630(53)}\\
\hline
{$P_1$}&{$24^3\times48$}&{6.21} &{3500} &{12}&{3}&{0.0667(5)} & {6.197(15)}\\
\hline
{$P_2$}&{$32^3\times64$}&{6.42} &{1958} &{20}&{4}&{0.0500(4)} & {11.270(38)}\\

\hline\hline
\end{tabular}
\caption{Simulation parameters for the HMC algorithm. $N_0$ is the number of integration steps, $\tau$ is 
the trajectory length and $t_{0}/a^2$ is the dimensionless reference Wilson flow time.}
\label{table1}
\end{center}  
\end{table} 
            
\section{Simulation details}
We have generated gauge configurations in SU(3) lattice gauge theory at 
different lattice volumes and gauge couplings using the {\tt openQCD} 
program \cite{openqcd}. Gauge configurations using periodic boundary conditions 
also have been generated for several of the same lattice parameters
(necessary changes to implement periodic boundary condition in 
temporal direction were made in the {\tt openQCD} 
package for pure Yang-Mills case). Details of the simulation parameters are
summarized in table \ref{table1}. In this table, $O$ and $P$ correspond to
open and periodic boundary configurations respectively.    

Topological susceptibility is measured over $N_{\rm cnfg}$ number of configurations
with two successive ones separated by $32$ thus making the total length of simulation time 
to be $N_{\rm cnfg}\times 32$.  
The lattice spacings quoted in table \ref{table1} are
determined using the results from Refs. \cite{gsw,necco}.
To smoothen the gauge configurations, Wilson flow \cite{wf1,wf2,wf3} 
is used and the reference flow time $t_0$ is determined 
through the implicit equation
\beq
\big\{t^2\langle\overline{E}(T/2)\rangle\big\}_{t=t_0} = 0.3
\eeq 
where $t$ is the Wilson flow time, $T$ is the temporal extent of the lattice 
and $\overline{E}$ is the time slice average of the action density given 
in Ref. \cite{open1}. Through this equation, the reference flow time provides 
a reference scale to calculate the physical quantities from lattice data.
An alternative to the $t_0$ scale is the $w_0$ scale proposed in Ref. \cite{wscale}.
We don't see any significant difference in our results using the two 
different scales.

\section{Numerical Results}
\begin{figure}[h]
\begin{center}
\includegraphics[width=5.5in,clip]
{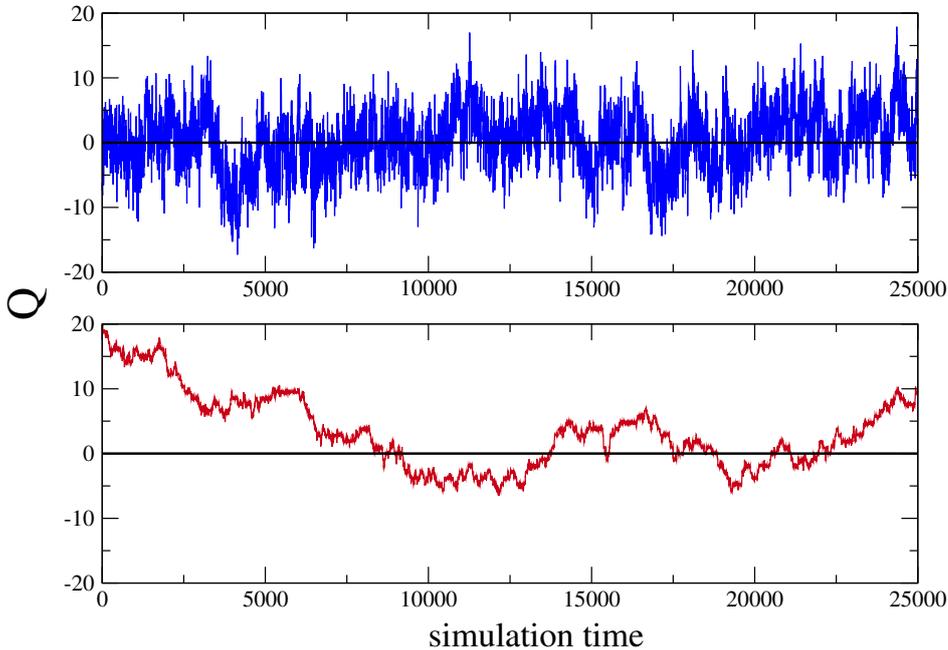}
\caption{Trajectory history of topological charge ($Q$) versus simulation time at $\beta=
6.59$ and lattice volume $48^3\times 96$ for open boundary condition (top) and 
 periodic boundary condition (bottom). The data shown is at Wilson flow time $t/a^2=2$. }
\label{history}
\end{center}
\end{figure}

\begin{figure}[h]
\begin{center}
\includegraphics[width=5.5in,clip]
{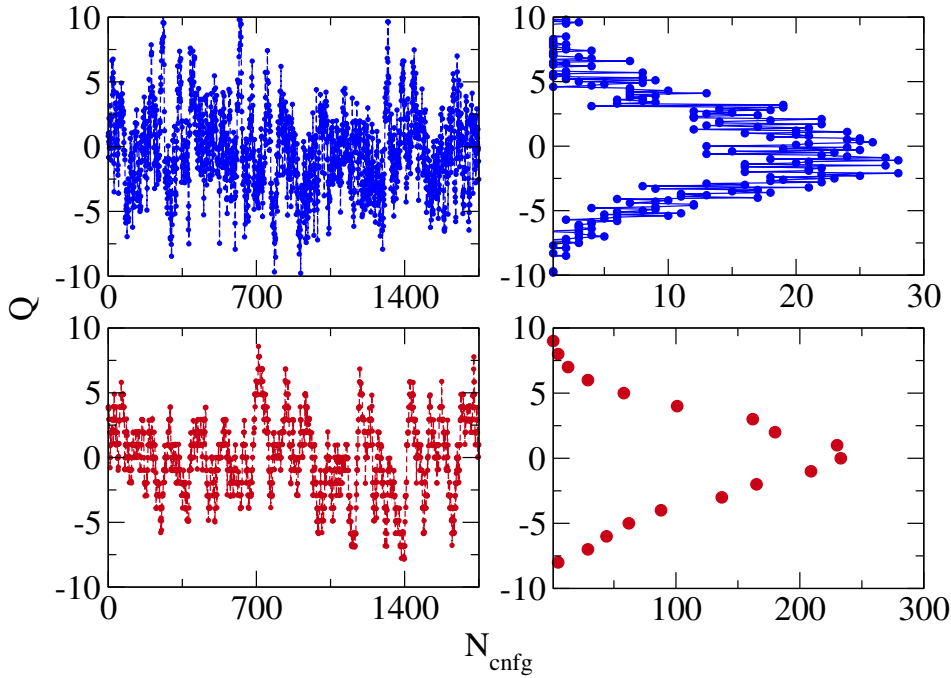}
\caption{Distribution of $Q$ versus $N_{cnfg}$. Top one (blue) is open boundary condition and bottom (red) is periodic
boundary condition at $\beta = 6.42 $ and lattice volume is $32^3 \times 64$. }
\label{Q_comp}
\end{center}
\end{figure}

\begin{figure}[h]
\begin{center}
\includegraphics[width=5.5in,clip]
{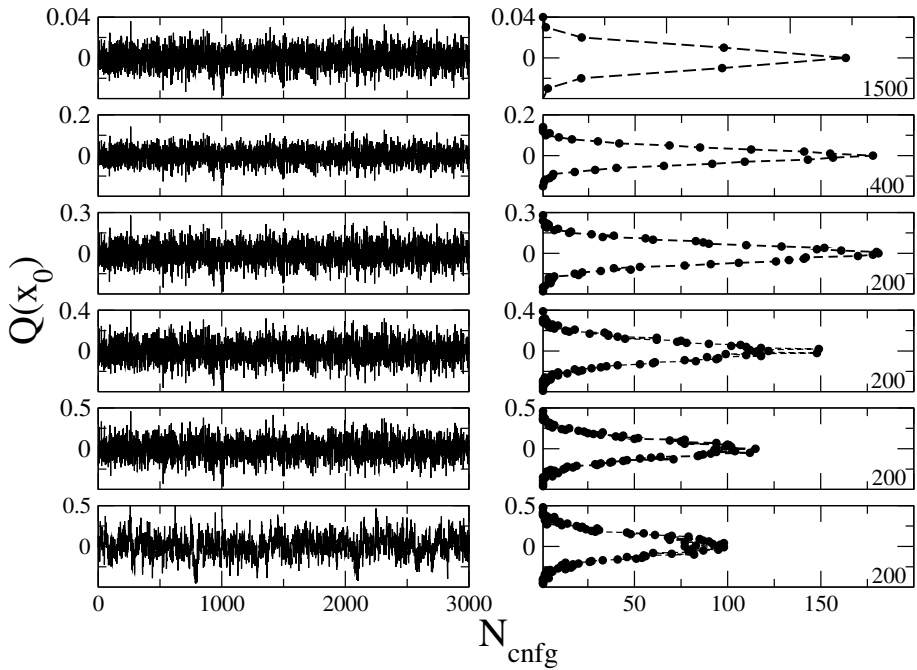}
\caption{Distribution of $Q(x_0)$ versus $N_{cnfg}$ for the ensemble $O_2$ where $x_0~ =~
0,~1,~2,~3,~4~{\rm and}~24$ from top to bottom respectively at $\beta = 6.42$
and lattice volume $32^3 \times 64$.  }
\label{Qslice}
\end{center}
\end{figure}

\begin{figure}[h]
\begin{center}
\includegraphics[width=5.5in,clip]
{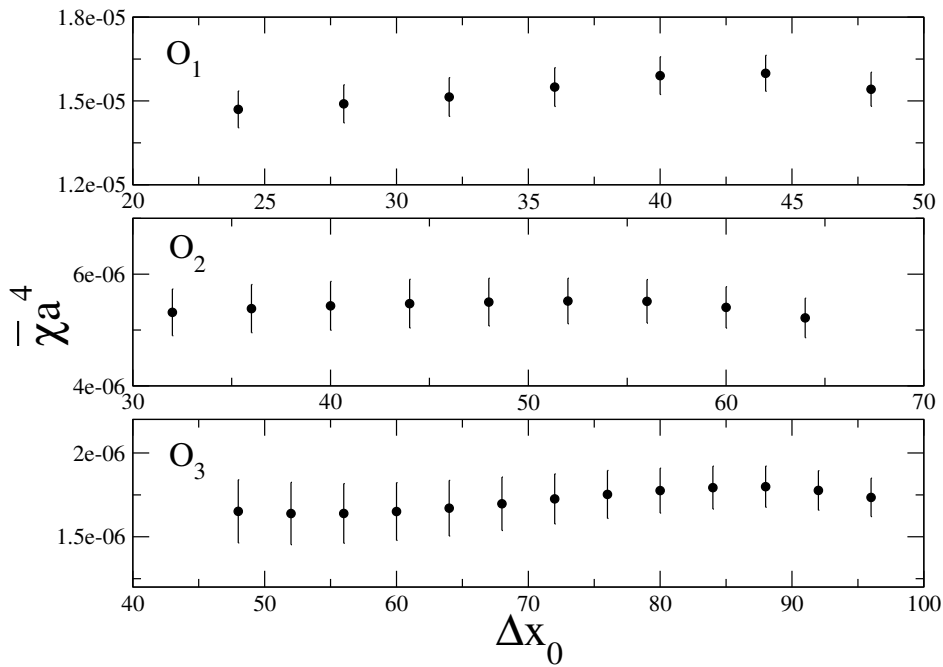}
\caption{Subvolume susceptibility ($\overline{\chi}$) versus temporal length ($\Delta x_0$)
for the ensembles $O_1$, $O_2$ and $O_3$.}
\label{subvol}
\end{center}
\end{figure}

\begin{figure}[h]
\begin{center}
\includegraphics[width=5.5in,clip]
{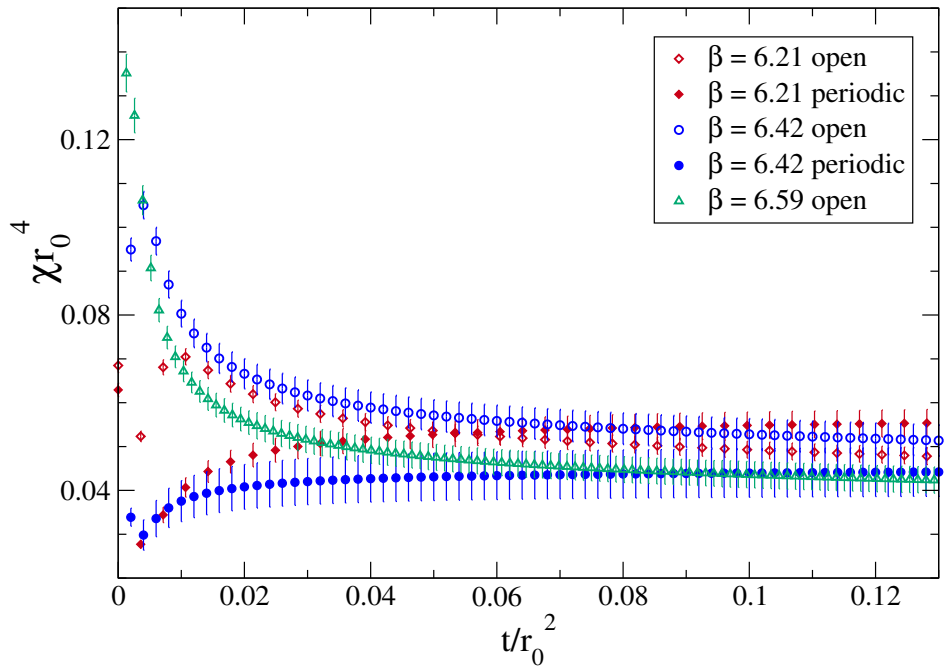}
\caption{Behaviour of topological susceptibility for both open and periodic
boundary condition under Wilson flow plotted versus the flow time for
different lattice spacings and lattice volumes. }
\label{wflow}
\end{center}
\end{figure}

\begin{figure}[h]
\begin{center}
\includegraphics[width=5.5in,clip]
{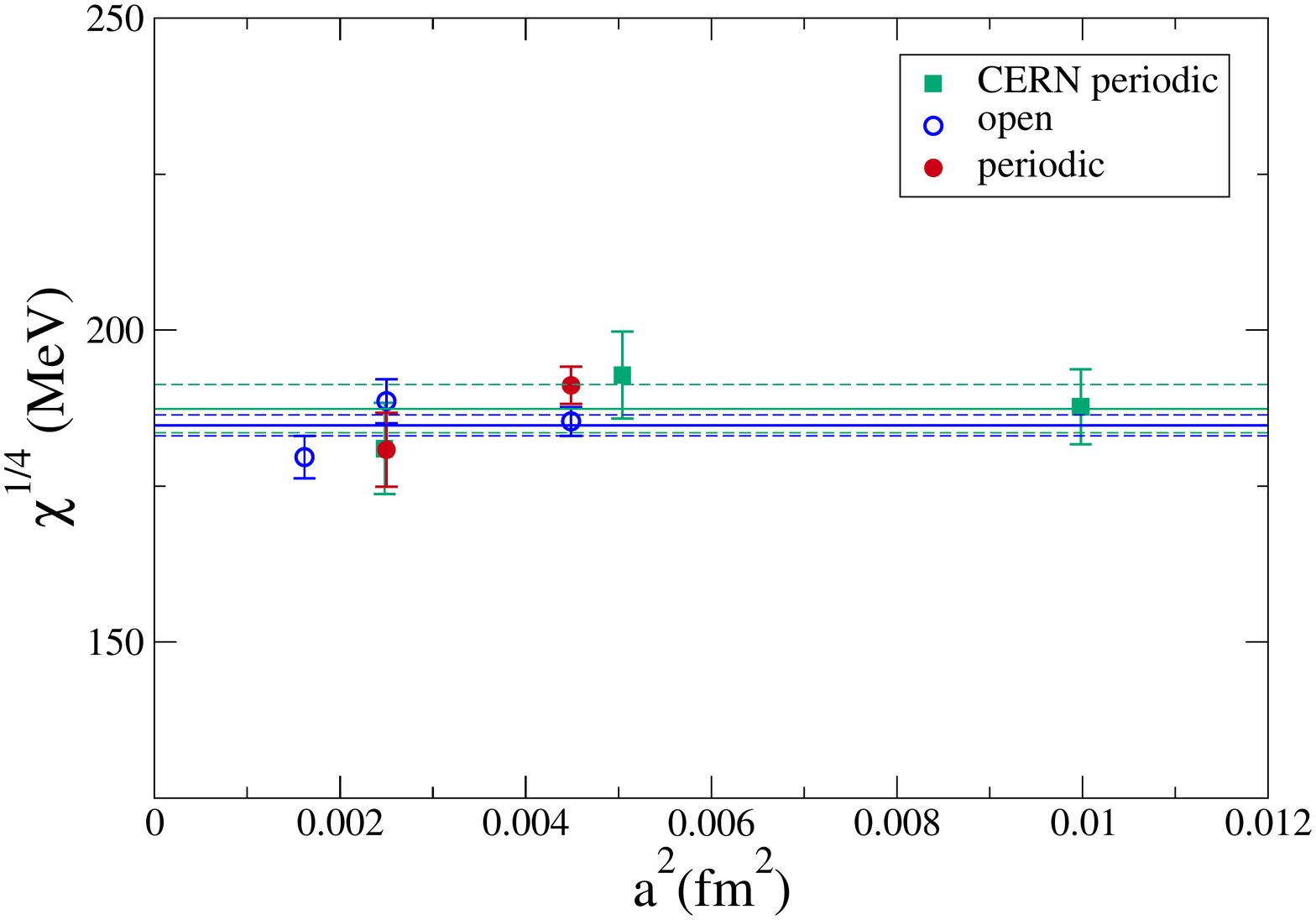}
\caption{$\chi^{1/4}$ in dimensionful unit plotted versus $a^2$ for both
open and periodic boundary condition for different lattice spacings and lattice
volumes. For comparison, data from Ref.\cite{lspb} for periodic boundary
condition is also plotted. Also shown are the linear fits to the data Ref.\cite{lspb} (green lines) and 
the data for open boundary condition (blue lines). }
\label{susflow}
\end{center}
\end{figure}

The open boundary condition has been proposed to help the tunneling of the system 
between different topological sectors characterized by the corresponding topological charge ($Q$)
as one approaches the continuum limit. To that end we first compare the trajectory history of $Q$ for
open versus periodic boundary conditions for a reasonably small lattice spacing. In figure
\ref{history} we plot the fluctuation of $Q$ versus simulation time at $\beta=
6.59 ~(a = 0.0402) $ and lattice volume $48^3\times 96$ for open boundary condition (top) and 
 periodic boundary condition (bottom) both starting from random configurations.
The data shown is at Wilson flow time $t/a^2=2$. Unless otherwise stated, all the data
presented in the following are at the reference Wilson flow time ($t_0$).
 It is evident 
that with open boundary condition, thermalization is reached very fast whereas with periodic
boundary condition it takes a long time just to reach thermalization. It is also evident that
after thermalization, autocorrelation length is much larger for the periodic boundary condition
compared to the open boundary condition. We have checked that the variation is not so marked for
periodic boundary conditions at larger lattice spacings.  
 
Next we look at the distribution of $Q$. In figure \ref{Q_comp} along with time histories,
we plot the histogram obtained for $Q$. Top one (blue) is open boundary condition and bottom (red) is periodic
boundary condition at $\beta = 6.42 $ and lattice volume is $32^3 \times 64$. We note that (1)
as expected from the boundary conditions, top (blue) $Q$ is not an integer whereas for bottom (red), it is an integer
and (2) even for this coupling ($\beta = 6.42$) which is lower compared to figure \ref{history}, taking
the same number of configurations, the top one gives much better spanning than the bottom.
In the plot of histograms in this figure, we have used bin sizes of $0.1$ (top) and $1$ (bottom).

One needs to investigate the effect of open boundary condition on topological charge density ($q(x)$).
We denote $q(x)$ integrated over the spatial volume at fixed Euclidean time $x_0$ by $Q(x_0)$.
The change in the behaviour of $Q(x_0)$ as a function of time slice $x_0$ reveals the effect of open boundary
in the temporal direction.  
The distribution of $Q(x_0)$ versus $N_{cnfg}$ is presented in figure
\ref{Qslice} for the ensemble $O_2$ where $x_0~ = ~0,~1,~2,~3,~4~{\rm and}~24$ from top to 
bottom respectively at $\beta = 6.42$ and lattice volume $32^3 \times 64$.
The distribution of $Q(x_0)$ is calculated with bin size of $0.01$.
 As we move from close to the 
boundary to deeper in the bulk, the spanning of $Q(x_0)$ steadily increases and finally settles down 
in the bulk region. The
same behaviour is also observed at the other end of the temporal lattice.

The topological susceptibility is defined as
\be
\chi = \frac{\langle Q^2 \rangle}{V}\nonumber
\ee
where $V$ is the space-time volume. To investigate the effect of open boundary on susceptibility 
we define a subvolume susceptibility \cite{forcrand} as follows:
\be
\overline{\chi}\left( \Delta x_0 \right) = \frac{\langle {\tilde Q}^2 \rangle}{{\tilde V}}\nonumber
\ee
where $ {\tilde Q} $  is the $q(x)$ integrated over the spatial volume and temporal length ($ \Delta x_0 $)
which is taken symmetrically over the mid point of the temporal direction. The subvolume ${\tilde V}$
is the product of spatial volume and $ \Delta x_0 $. In figure
\ref{subvol} we plot $\overline{\chi}$ versus $ \Delta x_0 $ for the ensembles $O_1$, $O_2$ and $O_3$.
Due to open boundary in the temporal direction, there is slight dip close to the temporal boundary 
which is consistent with the behaviour of $Q(x_0)$ as shown in figure \ref{Qslice}. 
We find that, overall, the effect of the open boundary on the subvolume susceptibility is within 
the statistical uncertainties.
  

It is interesting to study the stability of $\chi$ with respect to Wilson flow time.
In figure \ref{wflow}, we show the behaviour of $\chi$ for both open and periodic
boundary condition under Wilson flow plotted versus the flow time for
different lattice spacings and lattice volumes. For very early flow times, $\chi$ shows
non-monotonous behaviour for both
open and periodic boundary condition. For later flow times, $\chi$ converges from above to a
plateau for open boundary condition whereas it converges from below for the periodic boundary 
condition.
The values of susceptibility extracted at the reference flow time $t_0$ are 
given in table
\ref{table2} and plotted in figure \ref{susflow}. In the figure \ref{susflow}, 
we show $\chi^{1/4}$ in dimensionful 
unit plotted against $a^2$ for both open and periodic boundary condition for 
different lattice spacings and volumes. 
We find that the results for open and periodic lattices are very close to each other
at a given physical volume.

For comparison, data from Ref. \cite{lspb} for periodic boundary condition is also plotted. 
Also shown are the linear fits to the data Ref. \cite{lspb} (green lines) and the data for 
open boundary
condition (blue lines). The extracted value of $\chi^{1/4}$ for the open boundary condition data 
is 184.7 (1.7) MeV which
compares well with the result 187.4 (3.9) MeV of Ref. \cite{lspb}.  

\begin{table}
\begin{center}
\begin{tabular}{|c|l|l|}
\hline \hline
Lattice & $a^4\chi/10^{-5}$ &$\chi^{1/4}[{\rm MeV}]$\\
\hline\hline
{$O_1$}&{1.5418 (610)}&{185.4 (2.3)}\\
\hline
{$O_2$}&{0.5217 (354)}&{188.6 (3.5)}\\
\hline
{$O_3$}&{0.1794 (125)}&{179.6 (3.4)}\\
\hline
{$P_1$}&{1.7430 (973)}&{191.1 (3.0)}\\
\hline
{$P_2$}&{0.4407 (554)}&{180.8 (5.9)}\\
\hline\hline
\end{tabular}
\caption{Topological susceptibility.}
\label{table2}
\end{center}  
\end{table}

\section{Conclusions}

In this work we have shown that the open boundary condition in the temporal 
direction can
yield the expected value of the topological susceptibility in lattice 
SU(3) Yang-Mills theory.
The results agree with 
numerical simulations
employing periodic boundary condition. The advantage 
of open boundary conditions over
periodic boundary conditions (see, however, Ref. \cite{mm}) 
are illustrated in figure \ref{history}.

As further avenues of investigation, detailed comparison between Wilson flow
and conventional smearing techniques used for smoothening gauge fields
and the same between different algebraic as well as chirally improved fermionic
definitions of topological charge density are in progress. It is also interesting
to compute the topological charge density correlator (see Ref.\cite{topocorr} and the
references therein) using open boundary. 
\vskip .25in
{\bf Acknowledgements}
\vskip .1in
Numerical calculations are carried out on the Cray XT5 and Cray XE6 systems
supported by the 11th-12th Five Year Plan Projects of the Theory Division, 
SINP under the DAE, Govt. of India. We thank Richard Chang for the prompt 
maintenance of the systems and the help in data management. This work was in 
part based on the publicly available lattice gauge theory 
code {\tt openQCD} \cite{openqcd}.


\end{document}